\documentclass[a4paper]{spie}  

\usepackage{amsmath,amsfonts,amssymb}
\usepackage{graphicx}
\usepackage[colorlinks=true, allcolors=blue]{hyperref}

\usepackage{xcolor}

\title{ Testing deep learning techniques for event reconstruction in pair-production gamma-ray telescopes}

\author[ab]{Mattia Maniscalco}
\author[b]{Valentina Fioretti}
\author[b]{Nicolò Parmiggiani}
\author[b]{Andrea Bulgarelli}
\author[c]{Carolyn A. Kierans}
\author[c]{Adrien Laviron}
\author[b]{Gabriele Panebianco}
\author[a]{Alessio Aboudan}
\author[b]{Luca Castaldini}
\author[d]{Andreas Zoglauer}
\affil[a]{Università degli Studi di Padova, Via VII Febbraio , 2, 35122, Padova, Italy}
\affil[b]{INAF - Osservatorio di Astrofisica e Scienza dello spazio di Bologna, Via Gobetti, 93/3, 40129 , Bologna, Italy}
\affil[c]{NASA Goddard Space Flight Center, Greenbelt, MD 20771, USA}
\affil[d]{Space Sciences Laboratory, UC Berkeley, 7 Gauss Way, University of California, Berkeley, CA 94720, USA}

\authorinfo{Further author information: (Send correspondence to Mattia Maniscalco)\\Mattia Maniscalco: E-mail: mattia.maniscalco@phd.unipd.it, mattia.maniscalco@inaf.it}

\begin{document} 
\maketitle

\begin{abstract}
In the development of next-generation tracking detectors for gamma-ray space observations, the integration of advanced deep learning techniques into event reconstruction algorithms is a promising approach for improving the performance of the mission, reducing systematic uncertainties and maximizing the scientific output. 
In this study, we investigate the application of deep learning techniques such as Graph Neural Networks (GNNs) for the identification and reconstruction of particle tracks resulting from pair production events within the tracker of the AMEGO-X proposed concept. 
The goal is improving the angular resolution and the detection efficiency of the telescope, especially in the soft energy range (below 100 MeV) where multiple Coulomb scattering significantly degrades track reconstruction and consequently limits the sensitivity in the so-called MeV gap. 
We use the Geant4-based MEGAlib framework and a Python-based dedicated read-out and data handler processor to format the input datasets as close as possible to the real data. The simulated datasets are then used to train and evaluate two graph neural network architectures, GraphSAGE and Interaction Networks, and to compare their performance in pair-production event reconstruction in terms of Point Spread Function 68\% containment radius and effective area, also with that obtained using standard reconstruction techniques.
The results of this study indicate that graph neural network–based reconstruction is a promising approach for pair-production event reconstruction below 100 MeV when compared with standard reconstruction methods, with significant potential for further optimization.

\end{abstract}


\section{INTRODUCTION}
Pair-production tracking telescopes are the primary instruments used for gamma-ray astronomy at energies above tens of MeV, where pair production becomes the dominant interaction process. 
The silicon-tungsten trackers on board the ASI AGILE~\cite{TAVANI200852} (2007--2024) and NASA \textit{Fermi}~\cite{2009ApJ...697.1071A} (2008--present) satellites have played a fundamental role in high-energy astrophysics, providing major advances in the study of gamma-ray sources such as pulsars, active galactic nuclei, gamma-ray bursts, transient phenomena, and diffuse Galactic emission, while also contributing to the emergence of multimessenger astrophysics~\cite{2019NatRP...1..585M} through the observation of electromagnetic counterparts to gravitational-wave and neutrino events. Both AGILE-GRID~\cite{2009ApJ...697.1071A} and \textit{Fermi}-LAT~\cite{2009ApJ...697.1071A} use high-Z tungsten conversion foils to convert the photon into an electron-positron pair and silicon strip detectors to track the pair path, with an energy range from tens of MeV to hundreds of GeV. At low energies, the angular deflections induced by multiple Coulomb scattering (MCS), enhanced by the presence of high-Z passive material, make the reconstruction of the primary photon direction challenging. Different reconstruction techniques have been developed for these missions, primarily based on Kalman-filter tracking algorithms for particle trajectory reconstruction and vertex estimation. In the case of AGILE, dedicated reconstruction methods such as the AGILE REconstruction Method (AREM)~\cite{PITTORI2002295} and the on-board Kalman-filter-based tracking algorithm~\cite{GIULIANI2006692} were developed to improve the reconstruction of electron--positron pairs in silicon trackers. These approaches addressed key challenges such as three-dimensional track association, multiple-scattering effects, and the ambiguity in the identification of the pair-production plane, particularly for off-axis events. Similarly, the Fermi-LAT reconstruction chain~\cite{2013arXiv1303.3514A} is based on Kalman-filter techniques for track finding and fitting, combining information from the silicon-strip tracker and calorimeter in order to estimate the incoming photon direction and energy. In particular, the reconstruction algorithms are designed to account for multiple Coulomb scattering and for the complex topologies produced by pair-conversion events inside the tungsten converter foils. These techniques have been fundamental for achieving the high angular resolution and sensitivity required for modern gamma-ray astronomy. 
 
Below tens of MeV, the observational scenario is quite different. COMPTEL~\cite{1993ApJS...86..657S}, the last telescope to observe below 30 MeV and dismissed in 2000, had a sensitivity orders of magnitude worse than AGILE and Fermi. The Compton Spectrometer and Imager (COSI~\cite{tomsick2023comptonspectrometerimager}), a NASA small explorer mission to be launched in 2027, will cover the 0.2 - 5 MeV energy range by reconstructing Compton scattering. To unveil the entire soft gamma-ray band, missions such as ASTROGAM~\cite{Berge:2025kff} and AMEGO-X~\cite{Caputo_2022} proposed tracking detectors able to exploit both Compton scattering and pair-production by minimizing passive material while increasing the number of planes. However, MCS still dominates the track topology and new techniques, e.g. bayesian-based reconstruction~\cite{Aboudan_2022} for the ASTROGAM proposals, are being investigated to improve angular resolution and detection efficiency below 100 MeV.

 In recent years, machine-learning techniques based on Graph Neural Networks (GNNs) have emerged as particularly promising tools for charged-particle tracking applications, demonstrating significant potential in high energy physics experiments at particle colliders, such as those at the Large Hadron Collider (LHC~\cite{Caillou:2022hly,exatrkx_github,Duarte_2022}). GNNs naturally provide a graph representation of detector events. This approach is especially well suited for reconstructing the complex event topologies produced by MCS, which strongly affects low-energy electron--positron pairs in the MeV regime. The AstroPix~\cite{2023APS..APRN13009F, 2026NIMPA108170839S, STEINHEBEL2026171021} technology implemented in the AMEGO-X trackers is based on silicon high-voltage CMOS monolithic active pixel sensors (MAPS). These detectors provides a three-dimensional position information of the tracks , allowing the network to exploit the complete spatial topology of the event. This work represents one of the first applications of GNN techniques to pair-production event reconstruction in MeV gamma-ray telescopes, investigating advanced track reconstruction strategies using the AMEGO-X Tracker as a case study. The objective is to evaluate the potential of Graph Neural Network techniques to improve the angular resolution and sensitivity of next-generation MeV gamma-ray telescopes. This is achieved by simulating monochromatic gamma-ray photons with the AMEGO-X mass model using MEGAlib~\cite{2006NewAR..50..629Z} , a simulation framework based on Geant4~\cite{GEANT4:2002zbu} . Two GNNs architectures, GraphSAGE and Interaction Network, are investigated and their reconstruction performance is compared. The resulting angular resolution and effective area are then validated against the performance obtained with traditional reconstruction algorithms used in previous AMEGO-X studies.

The study begins with a description of the AMEGO-X detector mass model used for the simulations (Section~\ref{sec:geometry}). The two different Graph Neural Network architectures used are then presented and compared (Section~\ref{sec:gnn}). The simulation and reconstruction framework designed and adopted for the performance evaluation is subsequently described, including the graph-construction procedure, the tracking pipeline, and the event reconstruction strategy (Section~\ref{sec:sim_reco}). Finally, the obtained results are discussed in terms of preliminary values of PSF and effective area, providing a comparison with previous AMEGO-X performance studies based on classical reconstruction techniques (Section~\ref{sec:results}).

\section{THE AMEGO-X SPACE TELESCOPE}
\label{sec:geometry}
\begin{figure}[htbp]
\begin{center}

\begin{tabular}{cc}

\includegraphics[width=0.45\linewidth]{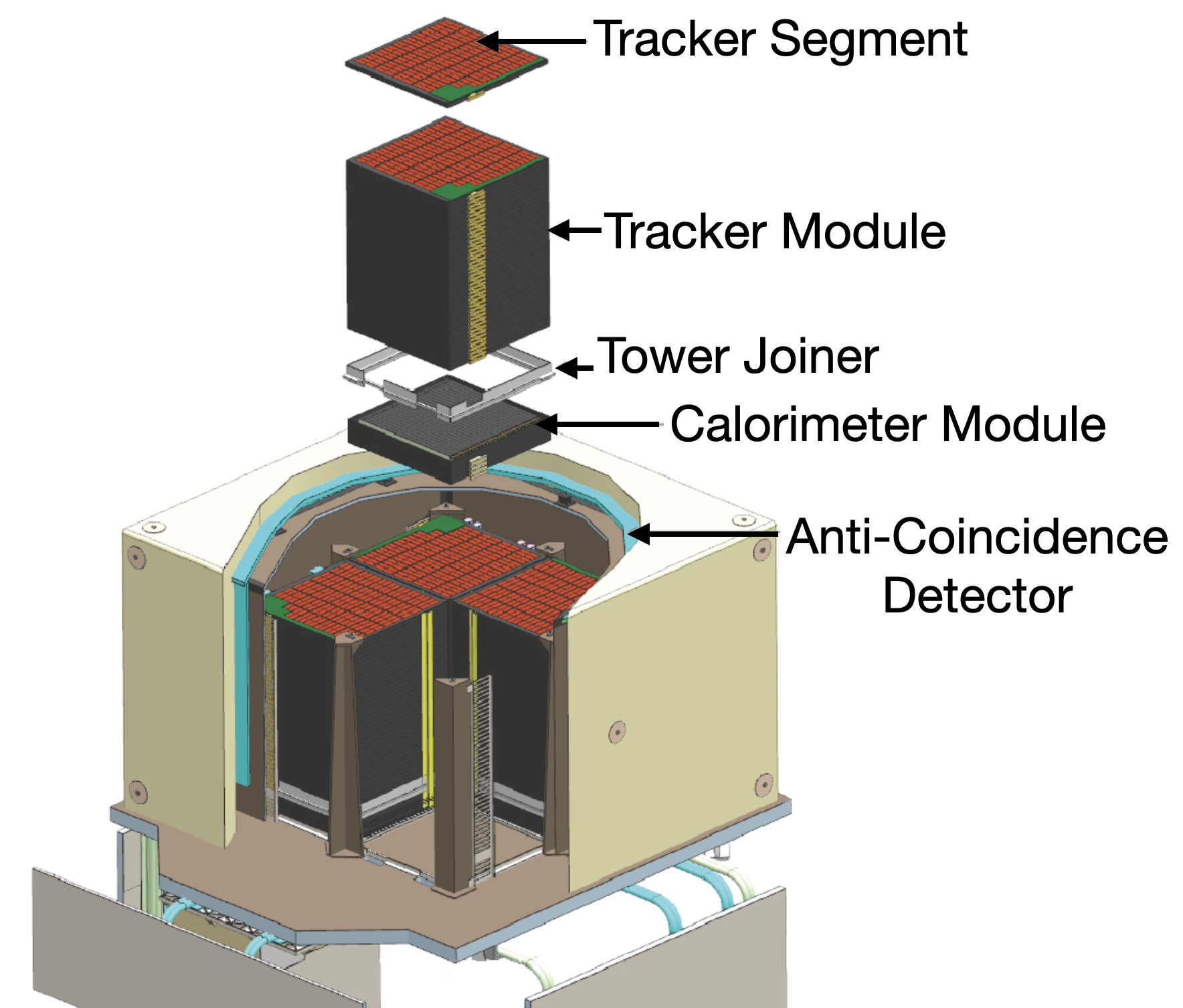} &
\includegraphics[width=0.45\linewidth]{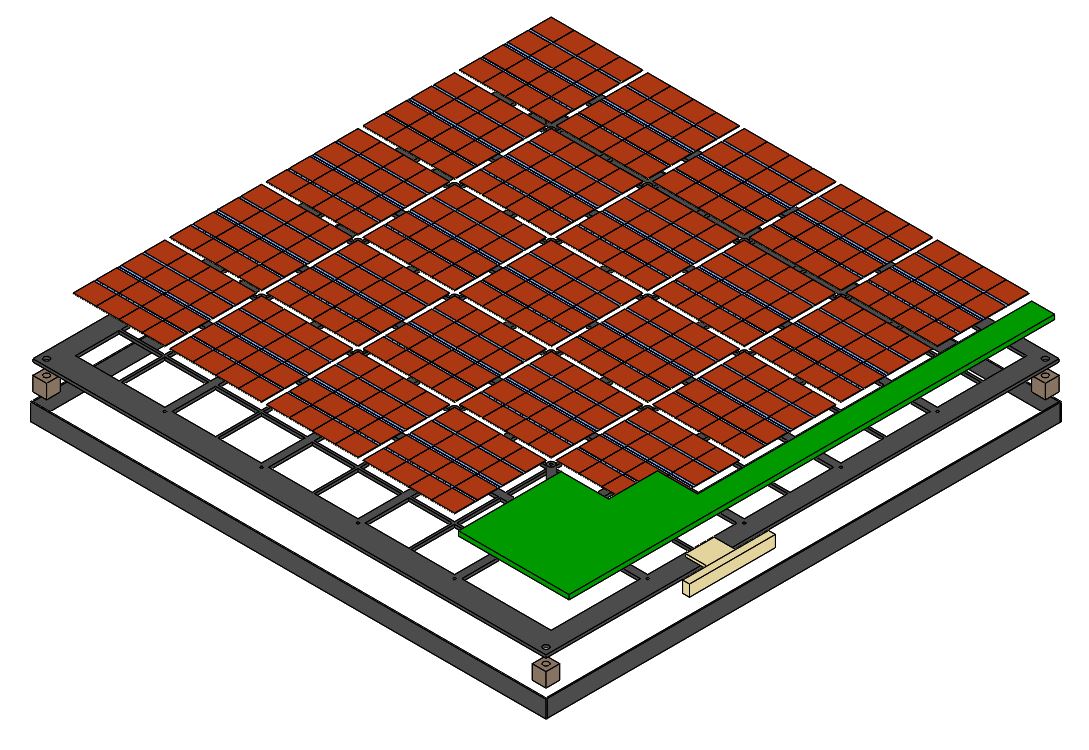} \\

{\footnotesize (a)} &
{\footnotesize (b)}

\end{tabular}

\end{center}

\caption{
\label{fig:amegox_detector}
(a): Overview of the AMEGO-X instrument. (b): schematic representation of the silicon Tracker structure\cite{Caputo_2022}.
}

\end{figure}

The All-sky Medium Energy Gamma-ray Observatory eXplorer (AMEGO-X)~\cite{Caputo_2022} is adopted as a case study for this work. AMEGO-X is a mission concept to be proposed to the next NASA Medium Explorer (MIDEX) program and designed to explore the poorly studied gamma-ray energy range between a few hundred keV and 1 GeV. AMEGO-X is conceived as a wide-field survey telescope capable of combining imaging, spectroscopy, and polarization measurements, enabling continuous monitoring of a large fraction of the sky and rapid detection of transient phenomena. The mission instrument consists of a Gamma-Ray Detector, composed of two main subsystems: a silicon Tracker and a Cesium Iodide calorimeter surrounded by an Anti-Coincidence Detector as depicted in Fig~\ref{fig:amegox_detector} (a). Together, they allow the detection and reconstruction of gamma rays through both Compton scattering at low energies and electron-positron pair-production at higher energies.

A key component of AMEGO-X is the pixelated silicon Tracker, shown in Fig~\ref{fig:amegox_detector} (b), which provides precise three-dimensional spatial and energy measurements of gamma-ray and charged-particle interactions. The Tracker is composed of four towers, each containing 40 stacked silicon detector layers separated by $1.5\,\mathrm{cm}$. Each layer consists of monolithic CMOS Active Pixel Sensors (APS), specifically developed for low-noise charged particle detection. The sensors, called AstroPix, are based on $0.5 \times 0.5\,\mathrm{mm}^2$ pixels integrated within fully depleted $0.5\,\mathrm{mm}$ thick silicon detectors. This technology combines excellent spatial resolution, low power consumption, and minimal passive material, which are crucial for accurately reconstructing low-energy pair-production events where multiple Coulomb scattering strongly affects the particle trajectories. The low passive material and high granularity of the AMEGO-X Tracker make it particularly well suited for advanced event reconstruction techniques, including graph-based and machine-learning approaches for track finding and vertex reconstruction thanks to the three-dimensional spatial measurements. 

\section{Graph Neural Networks Implementation}
\label{sec:gnn}

The Graph Neural Networks (GNNs)~\cite{4700287} are a class of \textit{Deep Learning} models~\cite{Goodfellow-et-al-2016} designed to operate on data represented as graphs. A graph is a mathematical structure composed of nodes and connections between them, referred to as edges. Unlike traditional neural network architectures, which are developed for regular data structures such as images or sequences, GNNs are specifically designed to process irregular and highly relational data. The core principle of GNNs is to associate each node with a feature representation that is iteratively updated by aggregating information from neighboring nodes. Through this \textit{message passing} mechanism, the network learns both the local properties of individual nodes and the global structure of the graph. As a result, geometric and topological relationships between the elements of the system are naturally incorporated into the learning process.

\begin{figure*}[ht]
    \centering
    \includegraphics[width=0.7\textwidth]{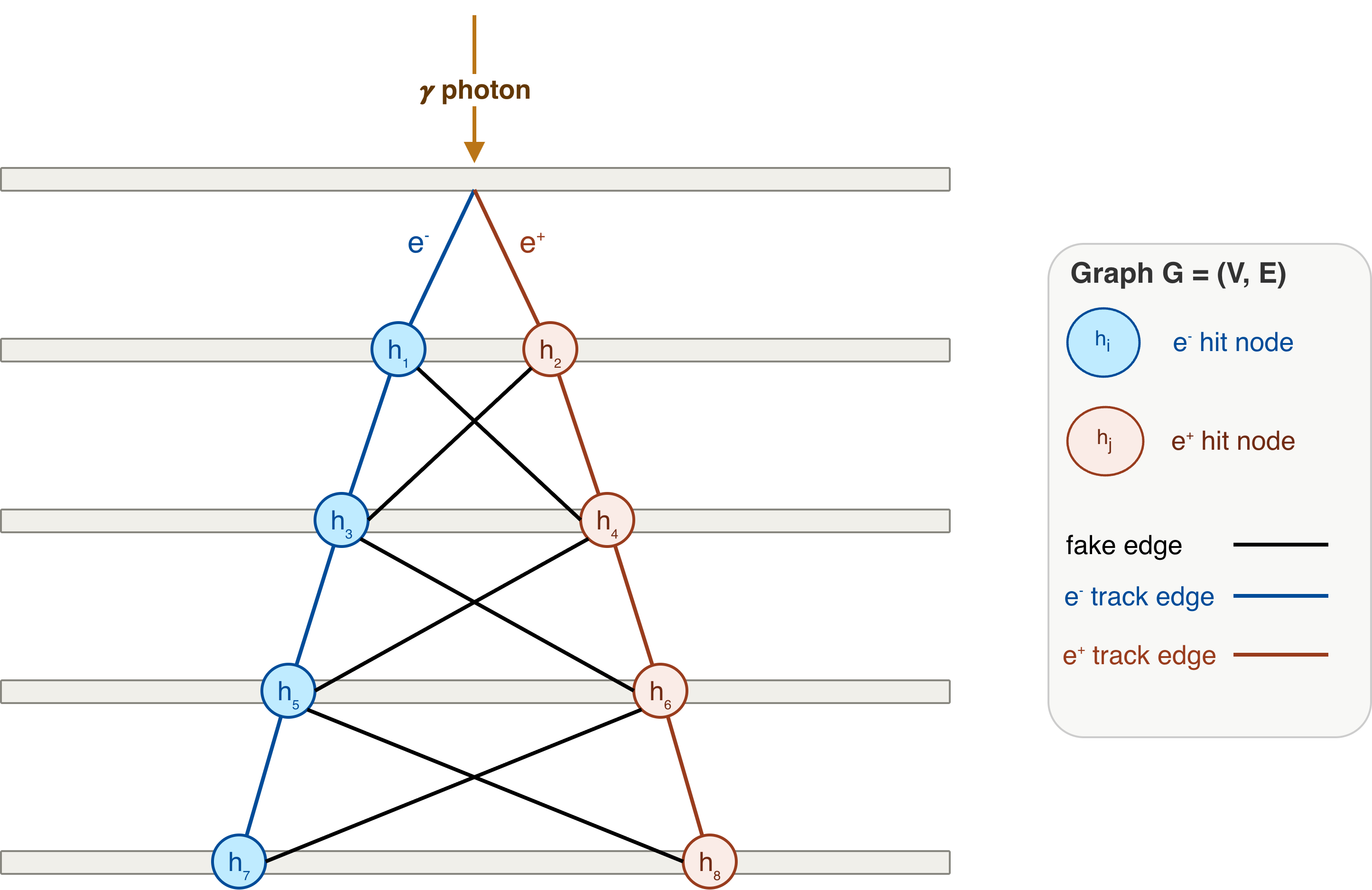}
    \caption{Schematic representation of the graph-based tracking approach used for pair-production event reconstruction. Detector hits are represented as nodes, while edges encode possible correlations between hits belonging to the same particle trajectory.}
    \label{fig:pair_prod}
\end{figure*}

In particle tracking and event reconstruction, GNNs are particularly promising because detector data can be naturally represented as graphs: detector hits correspond to nodes, while edges encode possible geometric or physical correlations between measurements, as depicted in Fig~\ref{fig:pair_prod}. This representation preserves the geometrical structure of the detector and allows the connectivity between hits belonging to the same track to be modeled directly. Compared to traditional tracking algorithms, such as Kalman filters~\cite{FRUHWIRTH1987444}, which often rely on combinatorial approaches or sequential reconstruction procedures, GNN-based methods provide greater flexibility in handling complex events, high detector occupancy, and irregular topologies, particularly in conditions where particle deflections are dominated by the non-Gaussian tails of the MCS, as is typical under hundreds of MeV. Furthermore, their ability to learn correlations directly from data makes them especially well suited for pattern recognition and track reconstruction tasks in modern tracking detectors.

In this case study, the purpose of the reconstruction task is to identify and reconstruct the two particle tracks generated by the electron--positron pair produced when an incoming photon undergoes a pair-production interaction in the tracker. The track reconstruction process can be divided into three main stages: \textit{track finding and track building}, \textit{track smoothing}, and \textit{track fitting}.  \textit{Track finding and building} identify and assemble detector hits into track candidates. \textit{Track smoothing} removes spurious hits and resolves ambiguities in the candidate trajectories. Finally, \textit{track fitting} estimates the track position, direction, and associated uncertainties through a dedicated fitting procedure. These steps are aimed at reconstructing the particle tracks and extracting the relevant information required to infer the properties of the primary photon responsible for the pair-production event. In particular, the GNNs approach presented in this work is employed during the \textit{track finding} and \textit{track building} stages, where correlations between detector hits are identified in order to reconstruct the two particle tracks generated by the pair-conversion process.

In this work, two different graph-based neural network architectures are investigated and compared. Both models operate at the edge-classification level, aiming to identify the connections between detector hits that are likely to belong to the same particle trajectory, following the graph-based tracking paradigm widely adopted in modern particle physics experiments in tracking pipelines~\cite{Caillou:2022hly,exatrkx_github,Duarte_2022}. The first architecture is based on an Interaction Network (IN) scheme~\cite{battaglia2016interactionnetworkslearningobjects}, originally designed to model relational interactions between objects through iterative edge and node updates and later extensively adopted in particle tracking applications. In this approach, both node and edge features are embedded into a latent space through dedicated encoder networks and iteratively updated through explicit edge-update and node-update blocks implementing a message-passing scheme. Node features are first projected into a latent embedding space through a multilayer encoder, while edge features are independently embedded using a dedicated edge encoder. At each message-passing layer, edge representations are dynamically refined using the concatenated latent features of the source node, destination node, and current edge embedding. Finally, the edge representation obtained during message passing is used,
together with the final node embeddings and the original edge features, to classify each candidate connection
through a multilayer perceptron\footnote{A multilayer perceptron (MLP) is a feed-forward neural network composed of fully connected layers and nonlinear activation functions, commonly used for classification tasks.
Nonlinear activation functions introduce nonlinearity into the network response, allowing the model to learn complex and nontrivial relationships between input features and target outputs.}. 

The second architecture is based on GraphSAGE~\cite{hamilton2018inductiverepresentationlearninglarge}, where node representations are updated by aggregating information from neighboring nodes. Input node features are first projected into a latent space through a linear embedding layer, after which successive \texttt{SAGEConv} layers iteratively update node representations by aggregating features from neighboring nodes using mean aggregation. After each layer, layer normalization, non-linear activation functions and dropout regularization are applied. Unlike the Interaction Network architecture, edge features are not dynamically updated during message passing. Instead, after the final node embeddings are obtained, the latent representations of the source and destination nodes are concatenated with the corresponding original edge features and passed to a multilayer perceptron performing binary edge classification. This comparison allows us to evaluate the impact of a more explicit edge-update mechanism, as implemented in the Interaction Network, with respect to a simpler neighborhood-aggregation approach based on GraphSAGE.

\section{Simulation and reconstruction pipeline}
\label{sec:sim_reco}

\begin{figure*}[ht]
    \centering
    \includegraphics[width=0.7\textwidth]{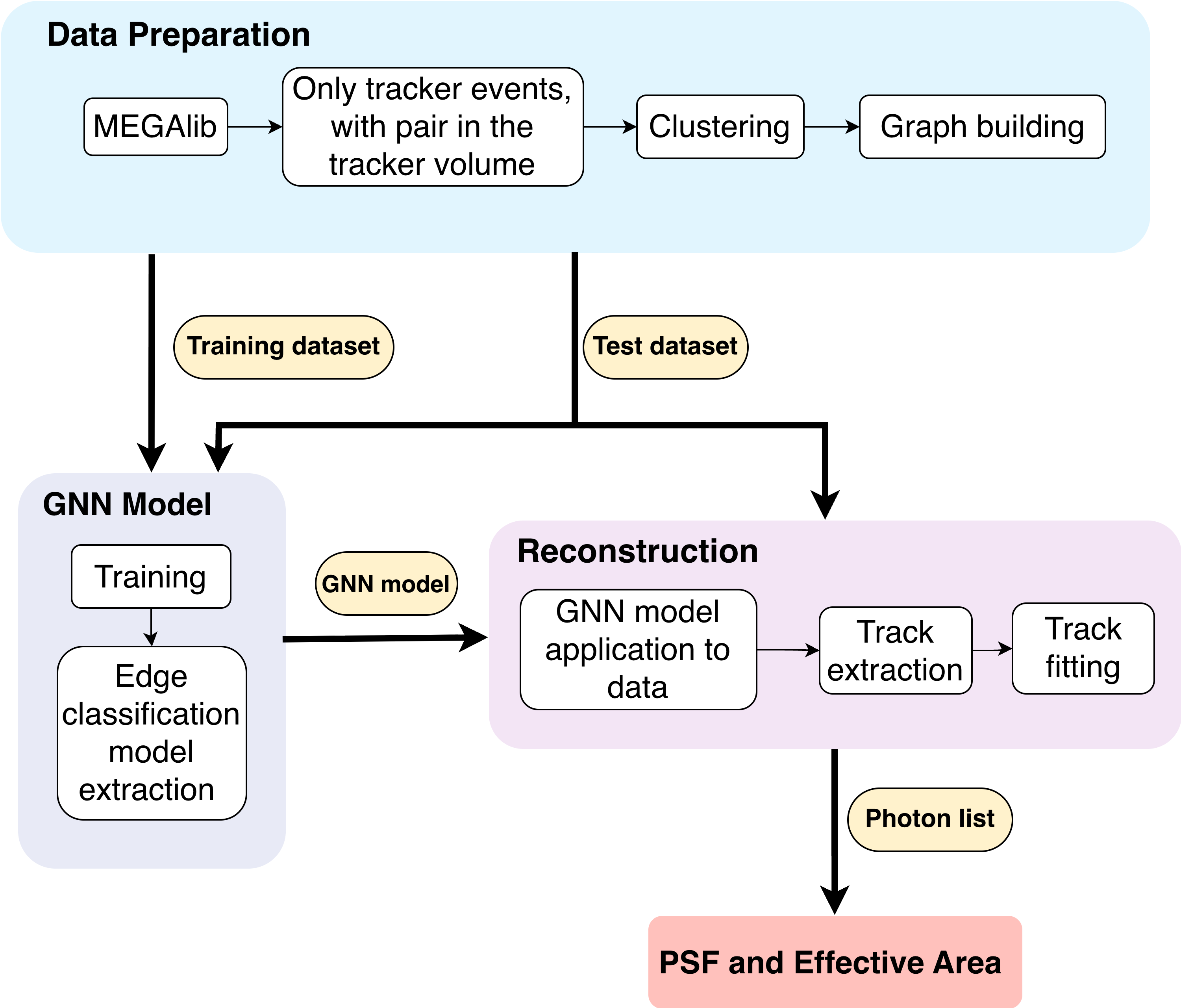}
    \caption{Schematic overview of the complete simulation, training and reconstruction pipeline used in this work. The workflow includes the MEGAlib simulation stage, graph construction, GNN training, track extraction and final PSF and effective area evaluation.}
    \label{fig:pipeline}
\end{figure*}

Fig.~\ref{fig:pipeline} summarizes the complete workflow adopted in this work, from the Monte Carlo simulation of pair-production events in the AMEGO-X tracker to the final reconstruction of the incident photon direction. The pipeline is organized into three main blocks: data preparation, graph-based neural network training, and final event reconstruction. Starting from simulated detector hits, clustered events are converted into graph representations suitable for GNN processing, where the network is trained to identify physically compatible hit connections belonging to the same particle trajectory. The reconstructed graph topology is subsequently used to extract the electron and positron tracks produced in the pair-conversion process and to estimate the incoming photon direction. The final reconstructed photon list is then employed for the evaluation of the effective area and a preliminary value of PSF in order to test the performance of the whole Deep Learning based reconstruction approach.

\subsection{Simulation and data preparation pipeline}

\begin{figure*}[ht]
    \centering
    \includegraphics[width=0.4\textwidth]{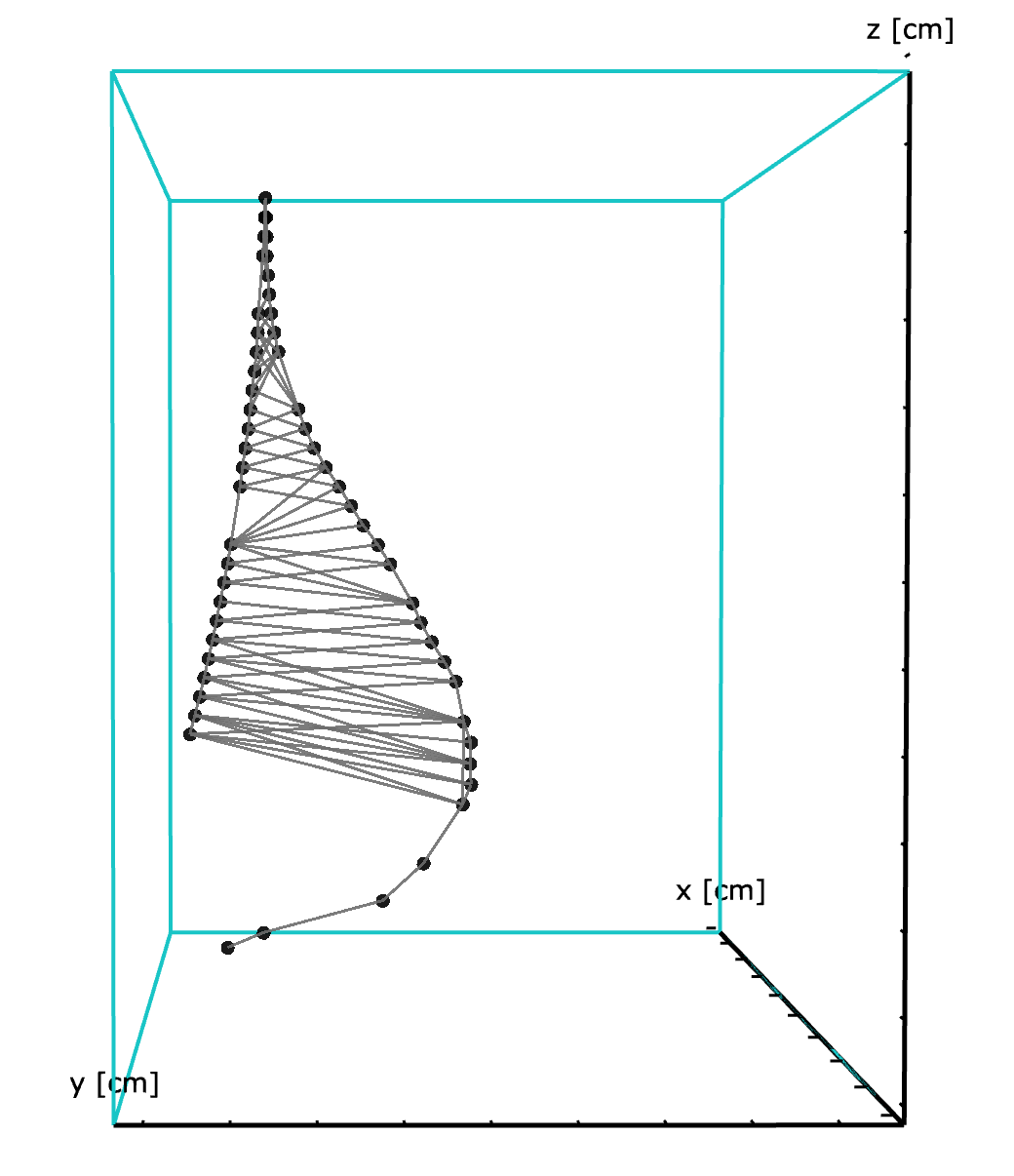} 
    \caption{Example of the graph generated from a clustered pair-conversion event at 100 MeV. The cyan lines in overlay are the border of the towers of the tracker. The black dots are the nodes of the graph that corresponds to the clustered detector hits, while the gray links between those points are the edges of the graph that represent the candidate track connections constructed according to geometrical and topological criteria.}
    \label{fig:graph_build}
\end{figure*}

The simulated dataset used in this work was generated using the \textit{MEGAlib}\cite{2006NewAR..50..629Z} simulation framework, which is built around \textit{Geant4}\cite{GEANT4:2002zbu}, adopting the complete mass model of the \textit{AMEGO-X}\cite{Caputo_2022} space telescope, including both active detector components and passive materials. The simulations were performed using gamma-ray beams at fixed energies (30, 50 and 100 MeV) and for two different photon incidence angles with respect to the detector normal. The incoming photons were generated using a far-field point-source configuration, corresponding to a parallel beam geometry in which all photons share the same incident direction and energy. This setup reproduces the observational conditions expected for astrophysical point sources located at infinite distance from the telescope. Only events corresponding to pair-production interactions occurring inside the tracker volume were considered. In particular, the event selection required the first physical interactions recorded in the simulation chain to be associated with the pair-production process, while the conversion point was constrained to lie inside the tracker fiducial volume. The simulated detector hits were then converted into ROOT~\cite{ROOT_NIMA_1997}-based data structures containing spatial coordinates, deposited energy, detector identifiers and Monte Carlo truth information. A dedicated clustering stage was subsequently applied to merge neighboring pixel hits belonging to the same particle track within each tracker plane. The clustering procedure exploited the detector geometry information directly available from the simulation output, including plane identifiers, tower identifiers and pixel coordinates. Adjacent pixels associated with the same Monte Carlo track were grouped together using a connectivity-based clustering algorithm, producing clustered hits characterized by energy-weighted spatial coordinates, cluster size and geometrical information of the pixels.
The clustering stage is adopted under the assumption of available clustering algorithms, capable of correctly identifying and separating the two particle trajectories generated near the pair-production conversion vertex. This assumption allows the study to focus specifically on the graph construction and track reconstruction performance of the proposed GNN-based approach, without introducing additional ambiguities related to low-level clustering inefficiencies or cluster merging effects. The clustered events were then converted into graph representations suitable for Graph Neural Network processing. Each cluster was represented as a node of the graph, while candidate edges were constructed by connecting clusters located on downstream tracker planes according to the expected physical propagation direction of the electron and positron tracks. The graph construction procedure intentionally employed an over-connected strategy, allowing forward connections across up to four detector planes. This choice was introduced to preserve track continuity in the presence of missing hits, for instance when particles cross passive material or detector regions without an active measurement. After the initial graph construction, a geometry-aware pruning procedure was applied to reduce nonphysical shortcut connections (e.g., edges connecting hits of the same track while skipping intermediate measurements that should belong to the reconstructed trajectory). This objective was achieved by removing multi-plane edges whenever an alternative path through intermediate active planes was available, while edges crossing only missing planes were preserved, since no intermediate measurement could support an alternative connection. Conversely, plausible bridge edges were also retained when they satisfied geometrical constraints on the total transverse displacement, the transverse displacement per crossed layer, and the three-dimensional distance. This procedure was designed to reduce geometrically inconsistent connections, which are particularly relevant in pair-production events where two nearby tracks are present. An example of a graph produced by the graph-building stage is shown in Fig.~\ref{fig:graph_build}.

\subsection{GNN Training}

The following section describes the training procedure adopted for the graph-based reconstruction. In the training stage, the problem is formulated as a binary edge-classification task on pre-constructed graphs. Each node represents a clustered detector hit, while candidate edges connect spatially compatible clusters located on downstream tracker planes. The GNN assigns a score to every candidate edge, corresponding to the probability that the two connected nodes belong to the same physical trajectory. Both architectures are trained as binary edge-classification models using the \texttt{Binary Cross Entropy with Logits} loss function. Given the strong class imbalance between true and false edges, a weighted loss is adopted through the introduction of a positive-class weighting factor (\texttt{pos\_weight}), computed from the ratio between negative and positive edges in the training dataset. This allows the models to penalize misclassified true connections more strongly and improves the reconstruction efficiency of valid track segments. The weighted Binary Cross Entropy with Logits loss is defined as

\begin{equation}
\mathcal{L} =
- \frac{1}{N}
\sum_{i=1}^{N}
\left[
w_p\, y_i \log\left(\sigma(x_i)\right)
+
(1-y_i)\log\left(1-\sigma(x_i)\right)
\right]
\end{equation}

where \(x_i\) represents the predicted logit for the \(i\)-th edge, \(y_i \in \{0,1\}\) is the corresponding ground-truth label, \(\sigma(x_i)\) is the sigmoid activation function, and \(w_p\) denotes the positive-class weighting factor. During training, an early stopping strategy was adopted in order to mitigate overfitting and improve the generalization capability of the models. The training procedure was monitored using the validation Average Precision (AP), defined as the area under the Precision--Recall curve, which provides a robust performance metric for imbalanced binary classification problems such as edge classification in graph-based tracking. The training was interrupted when no improvement in the validation AP was observed for 10 consecutive epochs, while the model parameters corresponding to the best validation performance were retained for the final evaluation. The networks output is a continuous score, referred to as a logit, for each candidate edge. Logits represent the raw, non-normalized outputs of the model before conversion into probabilities. During inference, the logits are transformed into probabilities through a sigmoid activation function. The final binary edge assignment is then obtained by applying a threshold on the predicted probabilities. The threshold was determined by selecting an optimal operating point on the validation set through maximization of the F1-score, which combines precision and recall into a single metric, computed from the precision--recall curve. This procedure provides a balanced compromise between reducing incorrectly reconstructed connections and preserving the completeness of the reconstructed tracks.

\subsection{Reconstruction pipeline}

\begin{figure*}[ht]
\begin{center}

\begin{tabular}{cc}

\includegraphics[width=0.4\linewidth]{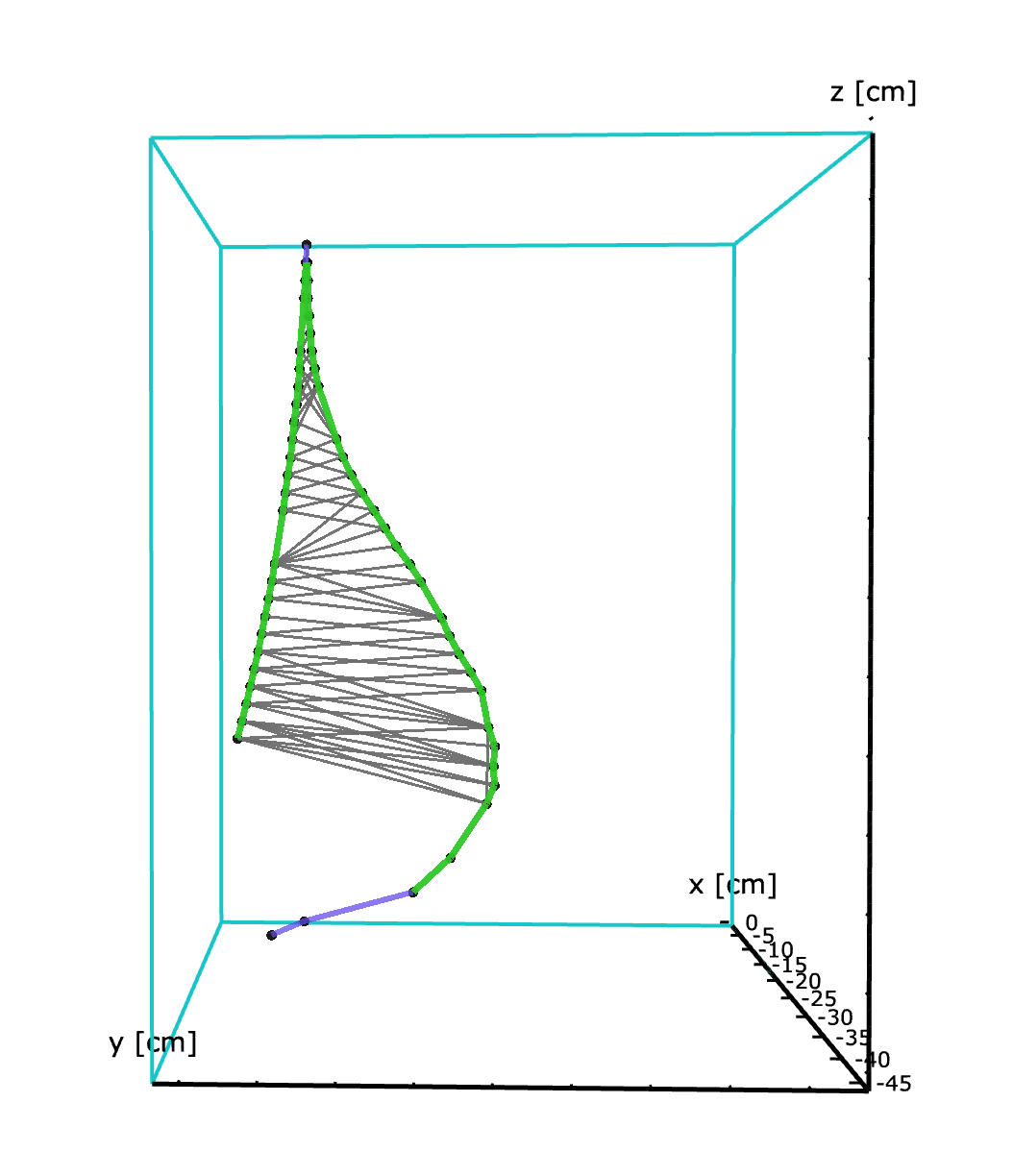} &
\includegraphics[width=0.385\linewidth]{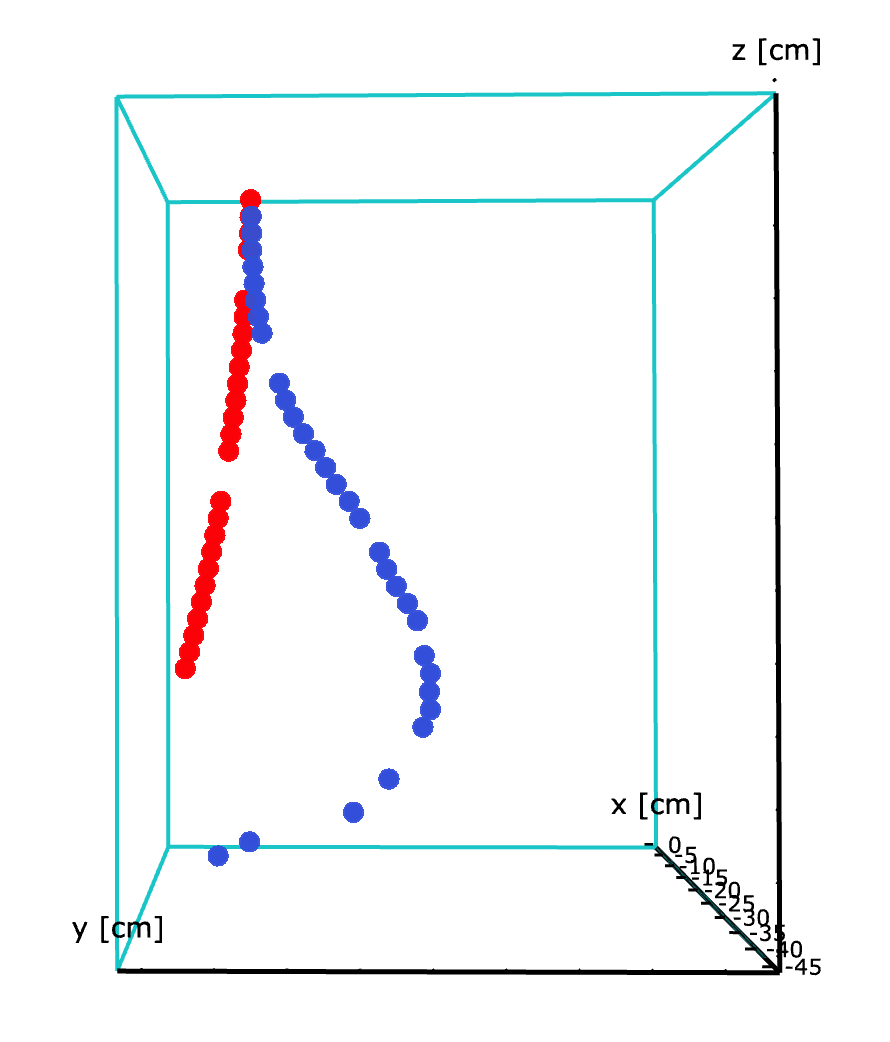} \\

{\footnotesize (a) GNN reconstruction} &
{\footnotesize (b) Monte Carlo truth}

\end{tabular}

\end{center}

\caption{
\label{fig:track_reco}
Example of the reconstructed graph for a pair-production event at 100 MeV. The cyan lines in overlay are the border of the towers of the tracker. Left: output of the GNN classification and track extraction procedure. The black dots represent the graph nodes corresponding to clustered detector hits. The green edges indicate the connections correctly classified by the GNN as belonging to the true tracks. Gray edges represent true-negative connections, correctly identified by the GNN as not belonging to the electron or positron trajectories. Violet edges correspond to false-negative connections, i.e. edges belonging to one of the true tracks but not identified by the GNN. Right: Monte Carlo truth information for the two reconstructed tracks, where the electron trajectory is shown in blue and the positron trajectory in red.
}
\end{figure*}

The reconstruction pipeline developed in this work, shown in Fig.~\ref{fig:pipeline}, consists of three main stages:

\begin{itemize}
    \item \textbf{Graph-based edge classification}, responsible for track finding and track building;
    
    \item \textbf{Track extraction}, corresponding to the track smoothing stage and aimed at identifying the two primary particle trajectories;
    
    \item \textbf{Track fitting}, used to reconstruct the incident photon direction from the extracted tracks.
\end{itemize}
The overall objective is the identification and reconstruction of the two particle trajectories generated by the electron and the positron produced during the pair-conversion process inside the detector. After the edge-classification stage, the reconstructed graph undergoes a dedicated track extraction procedure aimed at isolating the two dominant particle trajectories. After applying the score threshold, the graph is still not guaranteed to have a track-like structure. In fact, a single hit can remain connected to more than one downstream or upstream candidate hit, producing branches that are not compatible with the expected topology of a charged-particle trajectory.
For this reason, a pruning step is applied to resolve these ambiguities. The algorithm searches for nodes with multiple incoming or outgoing edges and keeps only the most consistent connection. The retained edge is not selected only from its GNN score, but also from the length and continuity of the chain that it supports. In practice, each competing edge is assigned a priority based on the number of connected segments that can be followed upstream and downstream from that edge. Edges belonging to longer and more coherent chains are therefore preferred, while isolated or locally inconsistent connections are removed. This procedure imposes an approximate one-in/one-out topology on the single node for what concern the edges, meaning that each hit can have at most one predecessor and one successor along the reconstructed trajectory. The output of this step is a simplified graph composed of track-like chains, which are then used as the starting point for the subsequent track elements connection and two-track extraction stages.
The resulting graph may still contain fragmented track segments caused by missing hits due to the passive material crossings or reconstruction inefficiencies. To address this issue, an additional post-processing stage introduces synthetic connections between disconnected track chains. Candidate reconnections are evaluated using geometrical compatibility criteria based on plane separation, transverse distance, local track direction consistency and linear extrapolation of the track trajectory. Additional merging procedures are also applied to reconnect track segments separated across detector tower boundaries or fragmented within the same tracker plane. An example of the output of the pair-track extraction and reconstruction procedure is shown in Fig.~\ref{fig:track_reco}, illustrating the performance of the GNN-based approach with respect to the Monte Carlo truth information. The green edges indicate the connections correctly classified by the GNN as belonging to the true electron and positron tracks.

Once the final connected graph is obtained, the two principal chains of hits are selected and identified as the pair tracks. A final fitting stage is then performed to estimate the incident photon direction. Each reconstructed trajectory is first globally fitted using independent linear fits in the XZ and YZ planes, from which an initial estimate of the pair-conversion vertex is obtained through the minimum-distance approach between the two fitted lines. An event selection criterion is then applied based on the distance of closest approach (DCA) between the reconstructed tracks, retaining only events with a DCA smaller than 0.5 cm. This requirement preferentially selects well-reconstructed events, improving the PSF, at the expense of a reduced event acceptance and therefore a lower effective area. In order to reduce the impact of multiple Coulombian scattering accumulated along the particle trajectories, a dedicated refit procedure is subsequently applied using only the hits closest to the reconstructed interaction vertex. The refitted track directions are then combined through a weighted vector sum in order to estimate the incoming photon momentum direction. 
Finally, the reconstructed direction is converted into the corresponding reconstructed source angular coordinates \((\theta,\phi)\)\footnote{The incoming photon direction is defined using the standard spherical coordinate convention. The polar angle $\theta$ is measured from the positive $z$-axis, while the azimuthal angle $\phi$ is measured in the $x$-$y$ plane starting from the positive $x$-axis and increasing toward the positive $y$-axis.}, which are used for the computation of the PSF. 

\section{Results}
\label{sec:results}

To evaluate the performance of the proposed track reconstruction pipeline, simulations were performed for three monochromatic photon-beam energies: 30 MeV, 50 MeV and 100 MeV. For each energy point, two different source configurations were considered: on-axis events and events generated at $\theta$=\(30^\circ\) off-axis in the \((\theta,\phi)\) plane with respect to the source direction. The simulated datasets were produced using the complete AMEGO-X mass model together with the \texttt{Cosima} simulation package within the MEGAlib framework. For each energy and inclination configuration, approximately \(7\times10^4\) pair-production events were generated and used for the preparation of the training datasets. After the event simulation stage, the events were converted into graph representations suitable for Graph Neural Network processing and subsequently used for model training. To evaluate the reconstruction performance, an independent large-scale test dataset containing \(10^6\) simulated events was generated for each energy configuration. Due to the energy dependence of both the pair-production cross section and the detector response, the final number of reconstructed pair-conversion events differs for each energy point, obtaining around 15000 events for 100 MeV to 4000 events for 30 MeV. The same graph-construction and pre-processing pipeline adopted for the training datasets was then applied to the test samples. During inference, the trained models were applied to the test graphs and the optimal edge-selection threshold determined from the validation set was used to perform the track reconstruction stages. The complete reconstruction chain was independently evaluated for both GNN architectures in order to compare their track reconstruction performance. The reconstructed photon directions were finally used to evaluate the angular resolution and detector response in terms of PSF and effective area. The effective area was computed following the formalism described in Zoglauer’s PhD thesis~\cite{2006PhDT.........3Z}:

\begin{equation}
A_{\mathrm{eff}} = A_{\mathrm{start}} \cdot \frac{N_{\mathrm{detected}}}{N_{\mathrm{started}}}
\end{equation}

where \(A_{\mathrm{start}}\) is the geometrical area of the generated source plane, \(N_{\mathrm{started}}\) is the total number of simulated events and \(N_{\mathrm{detected}}\) is the number of successfully reconstructed pair-conversion events. This allows a direct evaluation of the overall pair reconstruction performance of the proposed approach.

\begin{figure}[htbp]
\begin{center}
\begin{tabular}{c}
\includegraphics[width=0.85\linewidth]{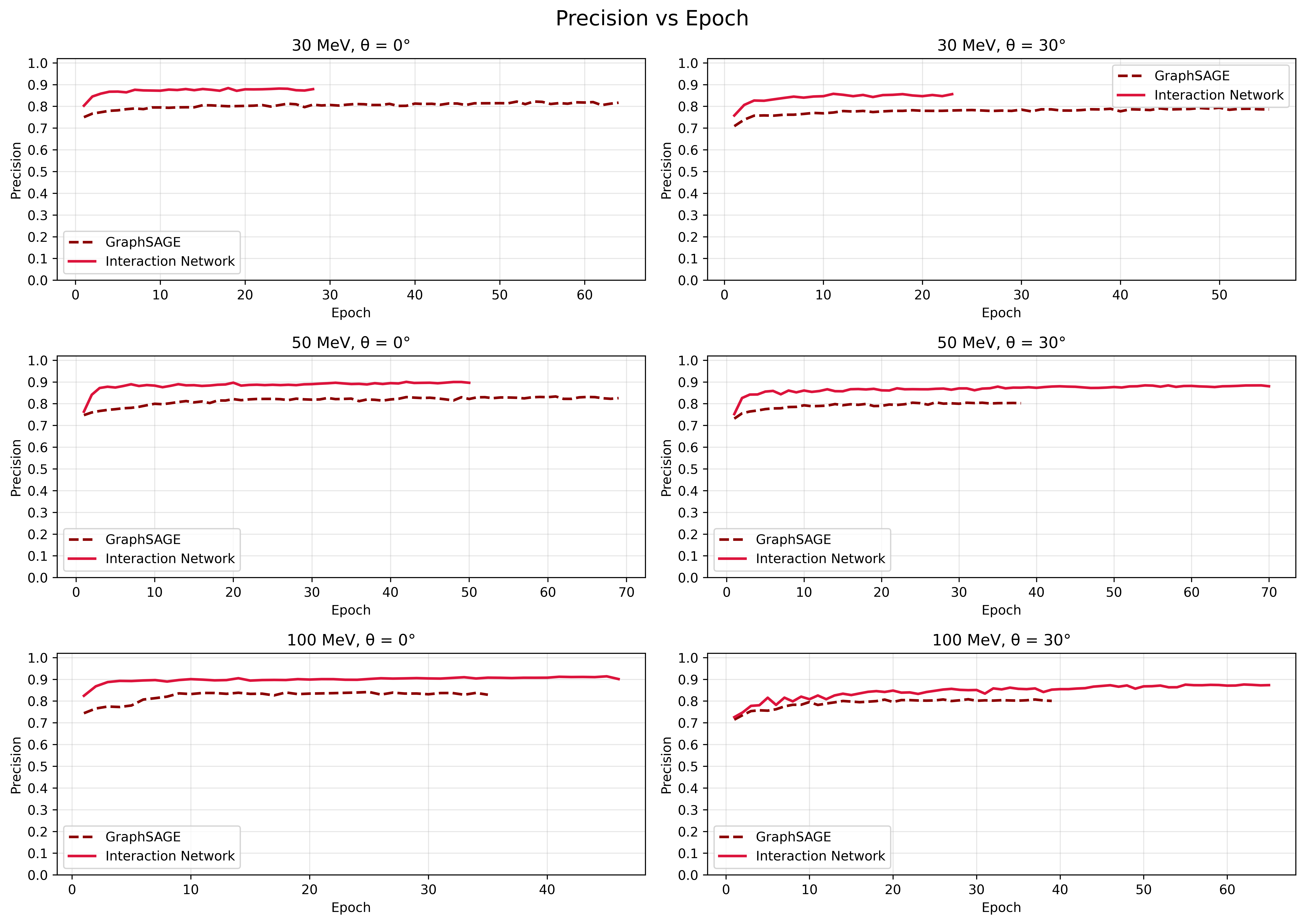}
\end{tabular}
\end{center}

\caption{
\label{fig:precision_vs_epoch}
Validation Precision as a function of the training epoch for the GraphSAGE (GS), dark red dashed line, and Interaction Network (IN),red line, architectures.
}
\end{figure}

\begin{figure}[htbp]
\begin{center}
\begin{tabular}{c}
\includegraphics[width=0.85\linewidth]{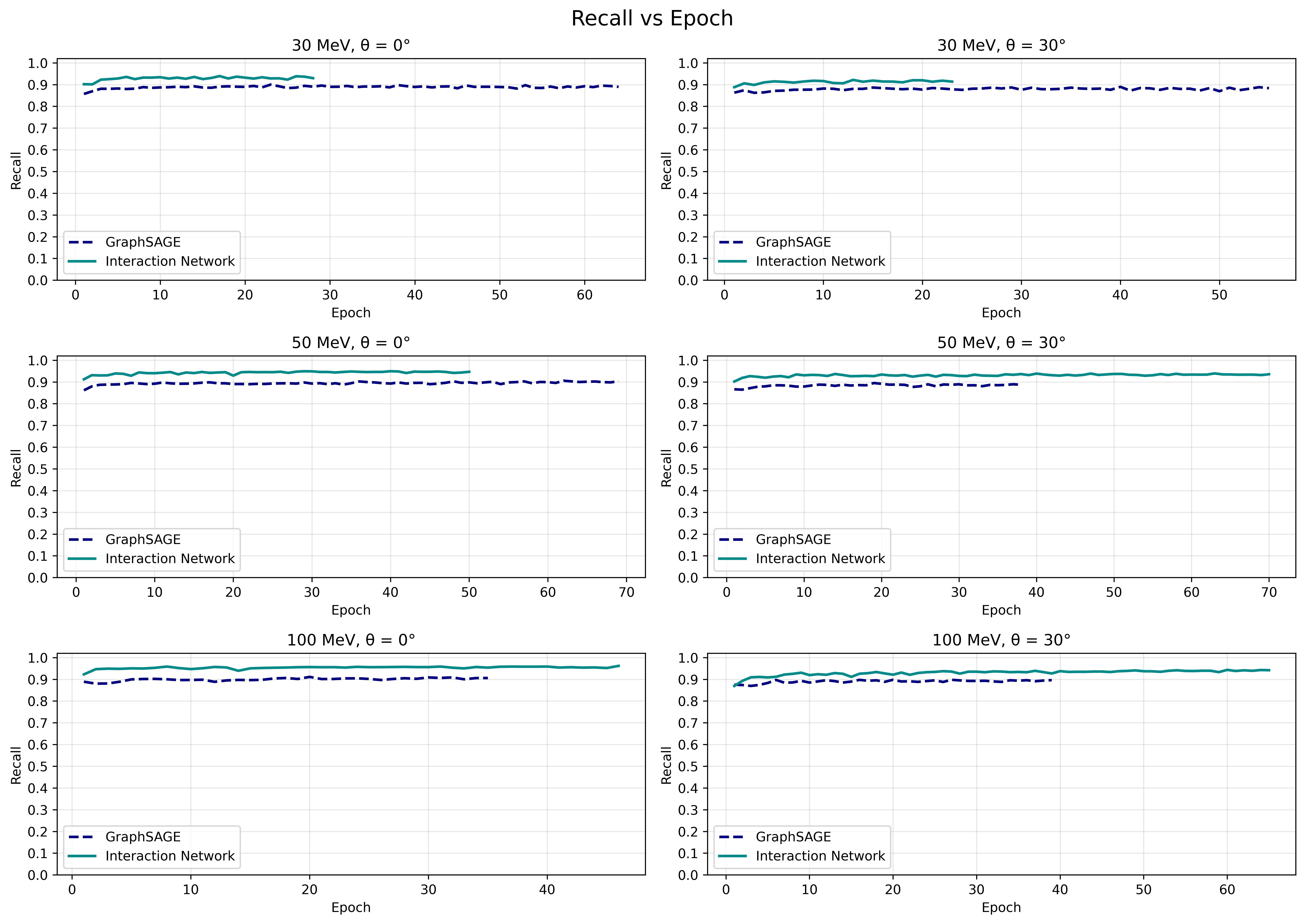}
\end{tabular}
\end{center}

\caption{
\label{fig:recall_vs_epoch}
Validation recall as a function of the training epoch for the GraphSAGE (GS), blue dashed line, and Interaction Network (IN), green line, architectures.
}
\end{figure}

To assess the reconstruction performance, the results are presented at two different levels: the network training performance and the performance of the complete reconstruction chain in terms of angular resolution and effective area. For the training stage, precision and recall were considered the most relevant metrics for this task, since they provide information about the capability of the network to correctly identify true track connections while preserving the completeness of the reconstructed trajectories.
The precision and recall metrics are defined as:

\begin{equation}
\mathrm{Precision} = 
\frac{TP}{TP + FP}
\label{eq:precision}
\end{equation}

\begin{equation}
\mathrm{Recall} = 
\frac{TP}{TP + FN}
\label{eq:recall}
\end{equation}

where \(TP\), \(FP\), and \(FN\) denote the number of true positive, false positive, and false negative edge classifications, respectively. In particular, a true positive (\(TP\)) corresponds to a candidate edge correctly identified by the network as belonging to the same physical particle track. A false positive (\(FP\)) represents an incorrectly reconstructed connection between hits that do not belong to the same track, while a false negative (\(FN\)) corresponds to a true physical connection that was either not identified by the network during the edge-classification stage or rejected by the final threshold selection applied to the assigned edge probabilities. 
Precision quantifies the fraction of reconstructed edges classified as true connections that actually belong to the same physical particle trajectory. A high precision therefore indicates a low contamination from incorrectly reconstructed connections. Recall instead measures the fraction of true physical connections that are successfully identified by the network, providing an estimate of the completeness of the reconstructed tracks. Fig.~\ref{fig:precision_vs_epoch} and Fig.~\ref{fig:recall_vs_epoch} shows respectively the precision and recall as a function of the training epoch for all energy configurations and source inclinations. Overall, the Interaction Network architecture achieves better performance than the GraphSAGE model, particularly for on-axis and higher-energy events. The best performance is obtained for the 100 MeV on-axis configuration, reaching a recall of approximately \(95\%\) and a precision close to \(90\%\). The reconstruction performance degrades for lower energies and off-axis events due to the stronger impact of multiple Coulomb scattering, which produces more complex event topologies and increases the ambiguities in the graph connectivity.

\begin{figure}[htbp]
\begin{center}

\begin{tabular}{cc}

\includegraphics[width=0.44\linewidth]{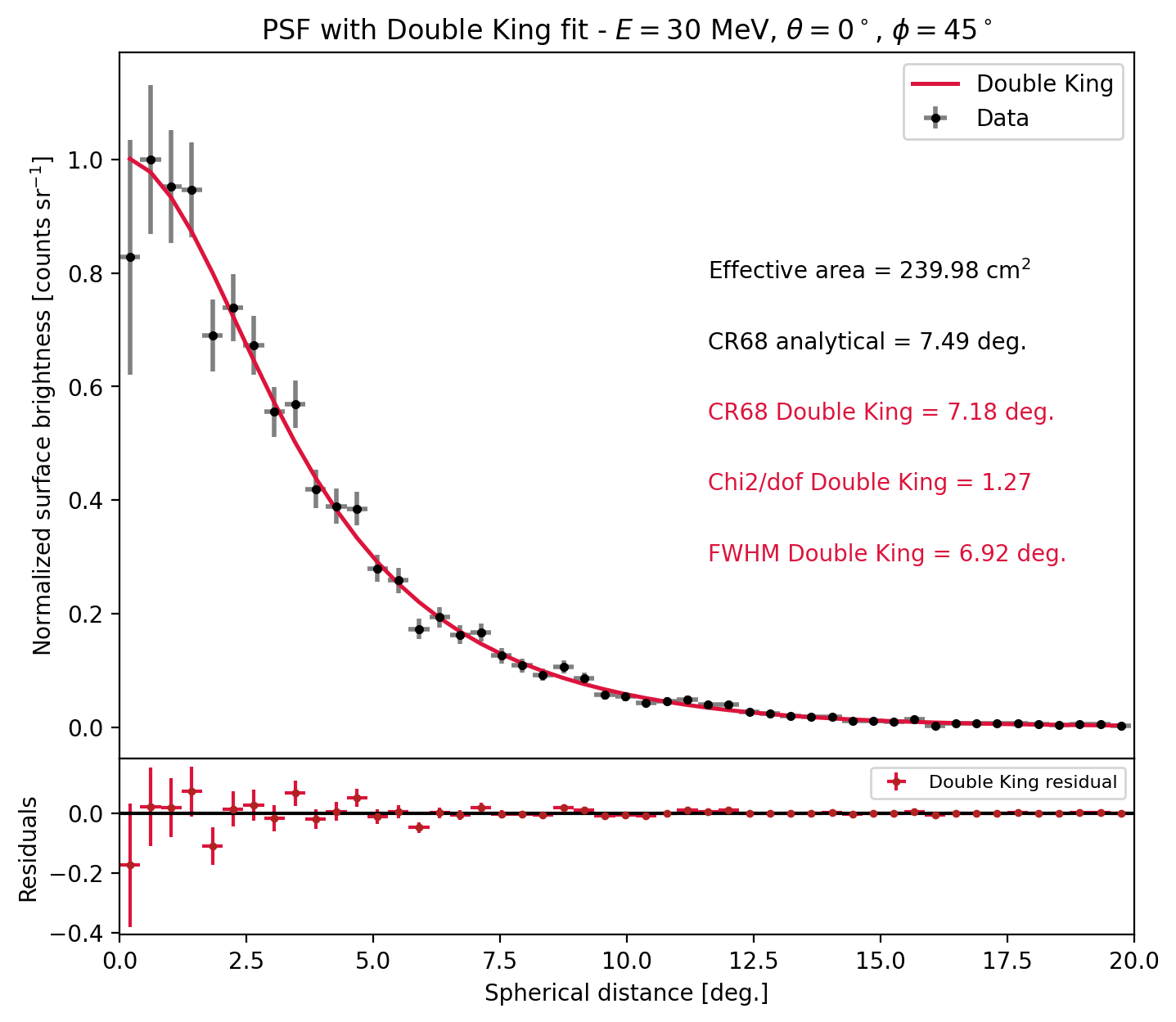} &
\includegraphics[width=0.44\linewidth]{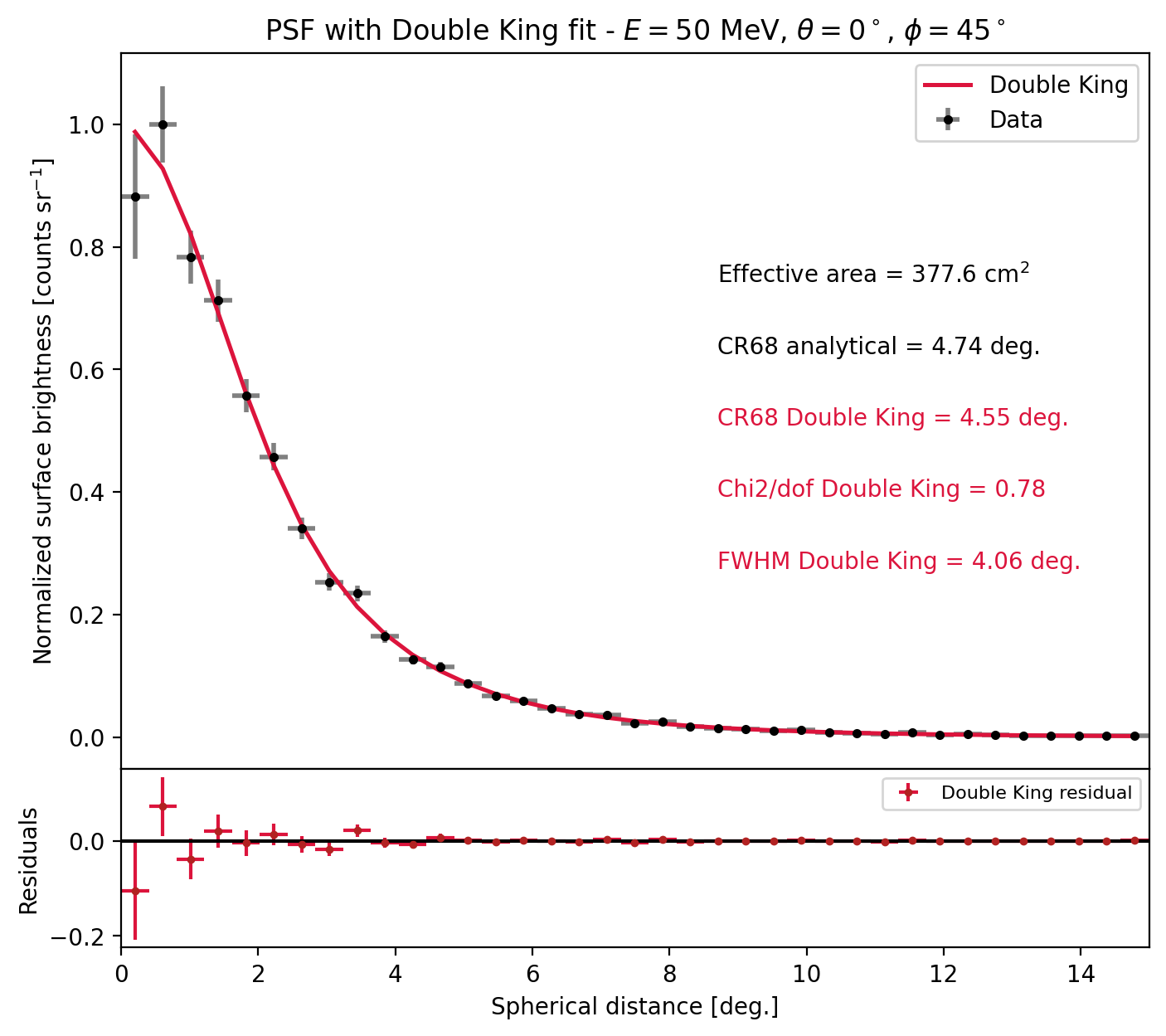} \\

{\footnotesize (a) 30 MeV} &
{\footnotesize (b) 50 MeV}

\end{tabular}

\vspace{0.2cm}
\includegraphics[width=0.44\linewidth]{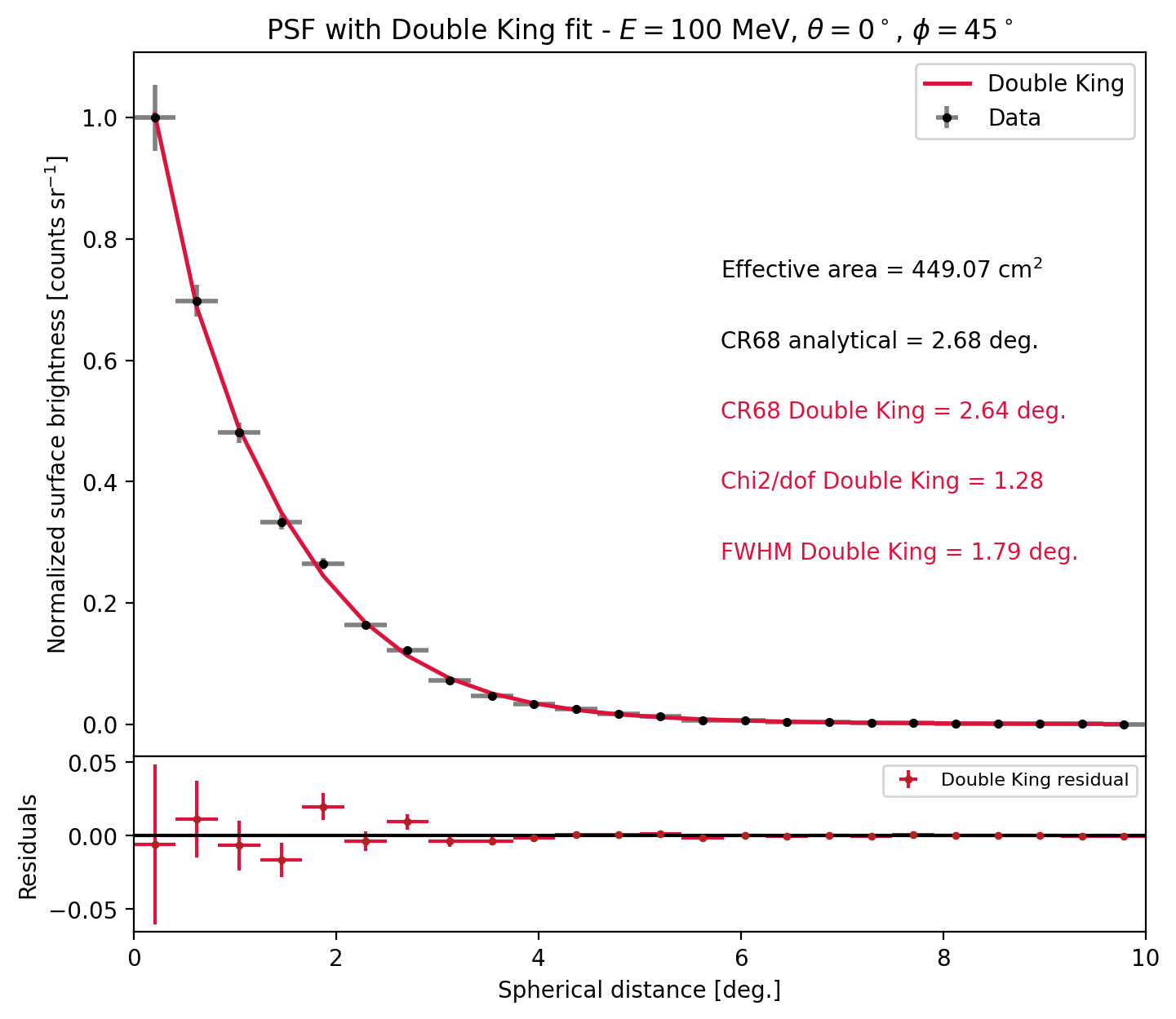}

{\footnotesize (c) 100 MeV}

\end{center}

\caption{
\label{fig:psf_king_compare}
Reconstructed PSF distributions obtained with the IN model at 30 MeV, 50 MeV, and 100 MeV. The red line corresponds to the Double King profile fit.
}

\end{figure}

The performance of the full reconstruction pipeline was subsequently evaluated through the computation of the PSF for each energy configuration and for both GNN architectures. The CR68 parameter corresponds to the angular radius containing \(68\%\) of the reconstructed events around the true source position. The PSF was first evaluated through the standard CR68 containment radius, obtained from the cumulative distribution of the event-by-event spherical angular distances between the reconstructed and true photon directions. In addition, the normalized surface-brightness profile was fitted using the same Double King function adopted in the FERMI-LAT analysis framework~\cite{ackermann_fermi_2012}, extracting the corresponding CR68 from the best-fit parameters. Examples of the reconstructed PSF distributions and the corresponding effective area for the 50 MeV and 100 MeV on-axis datasets using the Interaction Network architecture are shown in Fig.~\ref{fig:psf_king_compare}. 

A summary of the reconstructed PSF and effective area results is reported in Tab.~\ref{tab:psf_theta0} for on-axis events and in Tab.~\ref{tab:psf_theta30} for off-axis events with $\theta = 30^\circ$. As can be seen from the obtained results, the Interaction Network (IN) architecture systematically performs better not only during training, but also in the final reconstruction performance evaluation with respect to the GraphSAGE model. In particular, the IN approach provides consistently improved PSF values together with larger effective areas across all the considered energies and incidence angles. For on-axis events, the IN architecture improves the analytical CR68 from $7.74^\circ$ to $7.48^\circ$ at 30 MeV, from $5.21^\circ$ to $4.74^\circ$ at 50 MeV, and from $2.69^\circ$ to $2.67^\circ$ at 100 MeV, while simultaneously increasing the effective area from $215~\mathrm{cm}^2$ to $240~\mathrm{cm}^2$, from $342~\mathrm{cm}^2$ to $378~\mathrm{cm}^2$, and from $412~\mathrm{cm}^2$ to $449~\mathrm{cm}^2$, respectively. A similar behaviour is observed for off-axis events at $\theta = 30^\circ$, where the IN model improves the analytical CR68 from $9.46^\circ$ to $8.75^\circ$ at 30 MeV, from $5.99^\circ$ to $5.48^\circ$ at 50 MeV, and from $3.26^\circ$ to $3.08^\circ$ at 100 MeV. At the same time, the effective area increases from $148~\mathrm{cm}^2$ to $186~\mathrm{cm}^2$, from $328~\mathrm{cm}^2$ to $339~\mathrm{cm}^2$, and from $401~\mathrm{cm}^2$ to $411~\mathrm{cm}^2$, respectively.

\begin{table}[ht]
\caption{Comparison of PSF performance for the GraphSAGE and Interaction Network architectures at $\theta = 0^\circ$.}
\label{tab:psf_theta0}

\begin{center}
\resizebox{\linewidth}{!}{
\begin{tabular}{|c|cc|cc|cc|}
\hline

\rule[-1ex]{0pt}{3.5ex}
&
\multicolumn{2}{c|}{30 MeV} &
\multicolumn{2}{c|}{50 MeV} &
\multicolumn{2}{c|}{100 MeV} \\
\hline

\rule[-1ex]{0pt}{3.5ex}
GNN Architecture &
GraphSAGE & IN &
GraphSAGE & IN &
GraphSAGE & IN \\
\hline

\rule[-1ex]{0pt}{3.5ex}
Effective area [cm$^2$] &
214.53 & 239.98 &
342.19 & 377.60 &
411.89 & 449.07 \\
\hline

\rule[-1ex]{0pt}{3.5ex}
CR68 Analytical [deg] &
$7.74 \pm 0.64$ & $7.48 \pm 0.59$ &
$5.21 \pm 0.31$ & $4.74 \pm 0.29$ &
$2.69 \pm 0.16$ & $2.67 \pm 0.13$ \\
\hline

\rule[-1ex]{0pt}{3.5ex}
CR68 Double King [deg] (red.$\chi^2$) &
7.42 (1.09) & 7.18 (1.27) &
5.04 (1.07)& 4.67 (0.63)&
2.70 (1.32)& 2.68 (0.79) \\
\hline

\end{tabular}
}
\end{center}
\end{table}

\begin{table}[ht]
\caption{Comparison of PSF performance for the GraphSAGE and the Interaction Network architectures at $\theta = 30^\circ$.}
\label{tab:psf_theta30}

\begin{center}
\resizebox{\linewidth}{!}{
\begin{tabular}{|c|cc|cc|cc|}
\hline

\rule[-1ex]{0pt}{3.5ex}
&
\multicolumn{2}{c|}{30 MeV} &
\multicolumn{2}{c|}{50 MeV} &
\multicolumn{2}{c|}{100 MeV} \\
\hline

\rule[-1ex]{0pt}{3.5ex}
GNN Architecture &
GraphSAGE & IN &
GraphSAGE & IN &
GraphSAGE & IN \\
\hline

\rule[-1ex]{0pt}{3.5ex}
Effective area [cm$^2$] &
148.16 & 186.12 &
327.84 & 339.08 &
400.72 & 411.11 \\
\hline

\rule[-1ex]{0pt}{3.5ex}
CR68 Analytical [deg] &
$9.46 \pm 0.66$ & $8.75 \pm 0.57$ &
$5.99 \pm 0.39$ & $5.48 \pm 0.37$ &
$3.26 \pm 0.21$ & $3.08 \pm 0.18$ \\
\hline

\rule[-1ex]{0pt}{3.5ex}
CR68 Double King [deg] (red.$\chi^2$) &
8.79 (1.01) & 8.32 (0.99) &
5.81 (1.15) & 5.40 (1.31) &
3.22 (0.99) & 3.07 (0.96)\\
\hline

\end{tabular}
}
\end{center}
\end{table}

The obtained performance is finally compared with previous AMEGO-X pair-reconstruction results~\cite{Caputo_2022}, which were obtained using classical reconstruction algorithms. They report, for a on-axis point source:
\begin{itemize}
    \item At 100 MeV, a PSF of approximately $2.5^\circ$ with an effective area of about $400~\mathrm{cm}^2$.
    
    \item At 50 MeV, it reaches a PSF of approximately $3.5^\circ$ with an effective area of about $330~\mathrm{cm}^2$.
    
    \item At 30 MeV, it provides a PSF close to $5^\circ$ with an effective area of about $250~\mathrm{cm}^2$.
\end{itemize} 
Compared with the results obtained using the GNN-based reconstruction approach presented in this work, the classical reconstruction methods provide slightly better PSF values, whereas the GNN approach systematically achieves larger effective areas. This comparison shows that GNNs are indeed a promising reconstruction technique for pair-production telescopes operating at soft gamma-rays. It is also important to note that significant room for improvement remains in the GNN-based approach, particularly with regard to the training procedure and the subsequent track-fitting stage. Further optimization of these components is expected to improve the angular resolution while preserving the high reconstruction efficiency.

\section{Conclusions}
The obtained results appear promising, especially considering that several aspects of the reconstruction pipeline can still be significantly improved. In particular, better feature engineering and the introduction of more physics-informed approaches could further enhance the performance of the Graph Neural Network models.  Substantial improvements are also expected in the reconstruction stage following the network inference, especially in the track-extraction procedure and in the treatment of the network output. At the current stage, the fitting procedure is based on a relatively simple geometrical fit of the hit positions, without explicitly including physical information related to particle propagation and MCS. In contrast, traditional approaches based on Kalman-filter techniques model the effects of MCS during the track-fitting procedure, leading to a more physically consistent trajectory reconstruction. Therefore, implementing more advanced fitting strategies that include MCS modelling represents one of the most important future developments of this work. Future improvements will also focus on the optimization of the graph-construction procedure in order to explore lower-energy events, where the effects of multiple scattering become even more relevant and where the event topology becomes increasingly complex. In particular, low-energy reconstruction will likely require more sophisticated graph-building strategies and more robust tracking algorithms capable of handling highly scattered and fragmented trajectories. Overall, these results suggest that GNN-based approaches represent a promising direction for next-generation MeV gamma-ray event reconstruction and motivate further investigations in this field.


\bibliography{report} 
\bibliographystyle{spiebib} 

\end{document}